\documentclass[12pt]{iopart}
\usepackage{graphicx}
\usepackage{iopams}
\usepackage[pdfborder=000, colorlinks, linkcolor=blue, citecolor=blue]{hyperref}
\usepackage[numbers,sort&compress]{natbib}
\usepackage{color}

\usepackage{caption}
\begin{document}

\title{Phase transition and cascading collapse in binary decision-making dynamics}

\author{Xuyang Chen$^{1,2,3,5}$, Xin Wang$^{1,3,5}$, Longzhao Liu$^{1,2,3}$, Shaoting Tang$^{1,3,6}$ and Zhiming Zheng$^{1,3,4}$}

\address{$^1$ LMIB, NLSDE, BDBC, School of Mathematical Sciences, Beihang University, Beijing 100191, China}
\address{$^2$ ShenYuan Honor School, Beihang University, Beijing 100191, China}
\address{$^3$ PengCheng Laboratory, Shenzhen, 518055 , China}
\address{$^4$ Institute of Artificial Intelligence and Blockchain, Guangzhou university, Guangdong province 510006, China}
\address{$^5$ These authors contributed equally to this work}
\address{$^6$ Author to whom any correspondence should be addressed}

\ead{tangshaoting@buaa.edu.cn}

\vspace{10pt}

\begin{abstract}
Binary decision-making process is ubiquitous in social life and is of vital significance in many real-world issues, ranging from public health to political campaigns. While continuous opinion evolution independent of discrete choice behavior has been extensively studied, few works unveil how the group binary decision-making result is determined by the coupled dynamics of these two processes. To this end, we propose an agent-based model to study the collective behaviors of individual binary decision-making process through competitive opinion dynamics on social networks. Three key factors are considered: bounded confidence that describes the cognitive scope of the population, stubbornness level that characterizes the opinion updating speed, and the opinion strength that represents the asymmetry power or attractiveness of the two choices. We find that bounded confidence plays an important role in determining competing evolution results. As bounded confidence grows, population opinions experience polarization to consensus, leading to the emergence of phase transition from co-existence to winner-takes-all state under binary decisions. Of particular interest, we show how the combined effects of bounded confidence and asymmetry opinion strength may reverse the initial supportive advantage in competitive dynamics. Notably, our model qualitatively reproduces the important dynamical pattern during a brutal competition, namely, cascading collapse, as observed by real data. Finally and intriguingly, we find that individual cognitive heterogeneity can bring about randomness and unpredictability in binary decision-making process, leading to the emergence of indeterministic oscillation. Our results reveal how the diverse behavioral patterns of binary decision-making can be interpreted by the complicated interactions of the proposed elements, which provides important insights toward competitive dynamics.
\end{abstract}


\section{Introduction}
People often experience binary decision-making process when confronted with controversial issues or alternative options \cite{hu2017competing}. Integrated with complex social networks, the collective results at the population level emerging from such simple individual behavior have aroused great attention in many social problems \cite{barrat2008dynamical,castellano2009statistical,lazer2015rise}. For instance, the anti-vaccine movement \cite{wu2011imperfect,chen2019imperfect,johnson2020online}, the ceaseless spread of rumors and fake news \cite{grinberg2019fake,vosoughi2018spread}, the social polarization in U.S. presidential election \cite{barbera2015tweeting}, and etc. A natural way to understand the dynamics of binary decision-making is through the opinion evolution process. In the last decades, various mathematical models have been proposed to characterize the basic dynamical mechanisms that may rule the evolution of opinion formation \cite{degroot1974reaching,dandekar2013biased,becker2017network,noorazar2020recent,liu2020modeling}. Initially, opinions are assumed to diffuse in the form of discrete sets, such as the classic voter model \cite{clifford1973model,sood2008voter,fernandez2014voter}, Axelrod's model \cite{axelrod1997dissemination} and Sznajd model \cite{sznajd2000opinion}. Subsequently, a series of continuous-opinion models \cite{lorenz2007continuous}, where the opinions are described in terms of continuous numerical variables, are widely considered. Friedkin-Johnsen (FJ) model combines the effects of agent prejudices and social influence between connected neighbors \cite{friedkin1990social}. This framework has recently been extended to multidimensional situations where opinion evolves on several different topics \cite{parsegov2016novel}, and to belief interactions under logical constraints \cite{friedkin2016network}. Another important line of research is mostly based on the well-known Deffuant model \cite{deffuant2001mixing}, or the `bounded confidence model' (BCM), where the agents’ opinions would interact and converge to each other only if the difference of their opinions is smaller than a certain threshold. A further step is the introduction of network rewiring mechanism that drives the coevolution of networks and opinions \cite{holme2006nonequilibrium}, which is based on the fact that the network connections usually form only between individuals of similar ideas. Vicario \textit{et al.} find that the Rewire with Bounded Confidence Model and the Unbounded Confidence Model are able to reproduce the coexistence of two stable polarized opinions, while the Bounded Confidence Model not \cite{del2017modeling}. 

With the rapid development of online social platforms, the emerging empirical studies reveal the behavioral patterns of information consumption and opinion evolution, which significantly promotes the understanding of group decision-making \cite{schmidt2017anatomy,del2016spreading}. Recent advances largely focus on the widespread echo chamber phenomenon on large-scale social networks, where people form polarized homogeneous clusters of opinion \cite{cota2019quantifying,choi2020rumor}. The echo chambers can be interpreted as the combined effects of rewiring and peer influence under bounded confidence \cite{sasahara2019inevitability}, and can be particularly strong in more controversial topics \cite{baumann2020modeling}. Further, Wang \textit{et al.} take into account the conjugate influence of external political campaigns and online social influence process and identify the emergence of wavering clusters in addition to partisan echo chambers \cite{wang2020public}. 

Aside from these opinion evolution models, binary decision-making process has also been investigated via competitive spreading dynamics \cite{stanoev2014modeling,jie2016study}. Originated from the studies of competitive infectious diseases, the epidemic-like models, such as susceptible-infected-susceptible (SIS) model and susceptible-infected-recovered (SIR) model, are extended to study the competitive diffusion process \cite{karrer2011competing,sanz2014dynamics,leventhal2015evolution}. A recent work shows that the evolutionary advantage in competitive information diffusion could be changed or even reversed under the impacts of individual homogeneity trend \cite{liu2020homogeneity}. Furthermore, the complex contagion models that mimic the actual complicated mechanisms are also developed \cite{centola2007complex,stewart2019information}. 
 
While great efforts have been made on exploring the continuous opinion evolution as well as the discrete choice behavior, the coupled interactions of these two processes are ignored \cite{martins2008continuous,chowdhury2016continuous,varma2018continuous}. Consider the following case as an example: In a social innovation process, the innovation may provide higher payoff for its adopters and is therefore favored in public discourse, which drives the opinion dynamics \cite{shao2019evolutionary}. In turn, individual opinion decides decision-making behavior, which may reshape the macro-level choice distribution and determine whether the social innovation would succeed\cite{liu2021co}. Similar processes are widespread in competitive dynamics, especially in the reversal cases such as the competition between iWiW and Facebook in Hungary, calling for a thorough theoretical understanding \cite{torok2017cascading}. 

To this end, in this paper, we propose an agent-based model of binary decision-making process through competitive opinion dynamics. Our model incorporates three critical factors: bounded confidence, stubbornness level and opinion strength, within which bounded confidence and stubbornness level describe the continuous opinion evolution under social influence, while opinion strength characterizes the asymmetry influence of discrete choices on opinion dynamics. We show that the bounded confidence is the key determinant in giving rise to phase transition from co-existence to winner-takes-all state under binary decisions, corresponding to the changes from social polarization to consensus in public opinions \cite{prakash2012winner,proskurnikov2015opinion}. In particular, our model reveals that the initial advantage of a certain option could be reversed due to the combined influence of bounded confidence and asymmetry opinion strength. We also notice an interesting and counter-intuitive finding: an extremely strong opinion strength can instead weaken the competitive advantage in some parameter regions. Furthermore, we find the emergence of cascading collapse when the reversal happens, which is qualitatively the same as being observed in real data with regards to the failure of Gowalla \cite{leskovec2014snap}. Finally, we relax the homogenous assumption of population cognitive scope (i.e., bounded confidence) to a more realistic heterogenous situation. We highlight that the existence of individual cognitive heterogeneity can lead to the emergence of an indeterministic oscillation region, within which the final outcome of the binary decision-making process is unpredictable. Our model demonstrates how the coupled interplay between continuous opinion evolution and discrete choices brings about rich dynamical patterns and evolutionary outcomes in binary decision-making process, which might be applied to many competitive dynamics scenarios, such as marketing, boosting social innovations, and promoting non-pharmaceutical interventions during a pandemic \cite{glaubitz2020oscillatory, wang2020eco}.

\section{Theoretical framework}
\begin{figure*}[!h]
\includegraphics[width=0.95\textwidth]{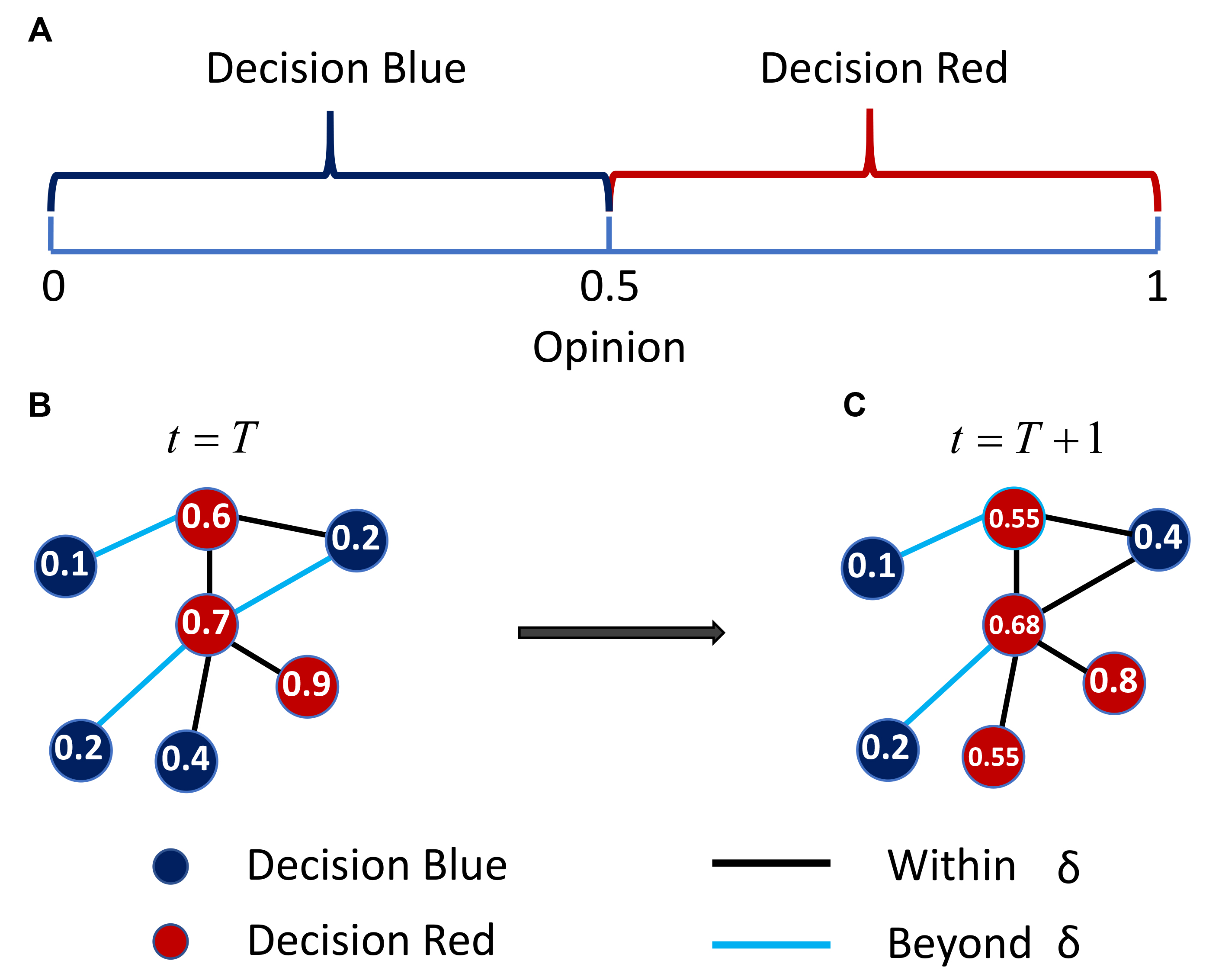}
\caption{Schematic of our dynamical model. (A) The function projecting continuous opinion space into discrete decision space. $x_i>0.5$ corresponds to Decision Red while $x_i<0.5$ corresponds to Decision Blue. (B)-(C) Dynamical processes of opinion evolution. We fix $\alpha=0.5$, $\delta=0.4$, $\beta=0.6$. The interactions take place along the black edges because the difference between the two endpoints' opinions are within bounded confidence $\delta$.} 
\label{Fig.1}
\end{figure*}

Consider a networked population with $N$ individuals, whose topological structure is recorded in the adjacent matrix $A=\{a_{ij}\}$. To be specific, if agent $i$ connects with agent $j$, $a_{ij}=1$. Otherwise, $a_{ij}=0$. 

In this paper, we mainly focus on binary-decision problems. Here the decisions are divided by Decision Red and Decision Blue. To further describe the individuals' preference for both decisions, we  introduce a continuous opinion space $[0,1]$. Let $x_i$ represent the opinion value of agent $i$. Naturally, the function transferring continuous opinion space to discrete decision space satisfies the following relationships (Fig. \ref{Fig.1}(A)). If $x_i<0.5$, agent $i$ would support Decision Blue. If $x_i>0.5$, agent $i$ would support Decision Red. Moreover, $x_i=0$ corresponds to the maximum preference for Decision Blue while $x_i=1$ means the maximum preference for Decision Red. 

Note that individuals' preference for different decisions belongs to category of opinions \cite{friedkin2016network}. Thus we could use modeling framework of opinion dynamics to describe the continuous evolution of individuals' preference (Fig. \ref{Fig.1}(B)-(C)). Confirmation bias, more likely to accept similar beliefs, has been recognized as a main mechanism reshaping opinion evolutions \cite{deffuant2001mixing}. This motivates us to consider bounded confidence $\delta$. Only when $|x_i-x_j|<\delta$, could agent $i$ and agent $j$ affect each other. 

In addition, we take into account the features of opinion themselves, i.e., the asymmetric persuasion, which is named by opinion strength. Without loss of generality, we set the opinion strength of Decision Red as $\beta$ and the opinion strength of Decision Blue as $(1-\beta)$. For each agent $i$, the set of neighbors who support Decision Red and have ability to affect agent $i$ can be written as
\begin{equation}
\Omega_{i}^{r}=\{j|a_{ij}=1, |x_i-x_j|<\delta,x_i>0.5\} 
\end{equation} 
Similarly, 
\begin{equation}
\Omega_{i}^{b}=\{j|a_{ij}=1, |x_i-x_j|<\delta,x_i<0.5\} 
\end{equation}

Integrating bounded confidence and opinion strength, at each time step, the updating process of $i$'s opinion from $x_i$ to $\tilde{x_i}$ satisfies the following rule:
\begin{equation}
\tilde{x_i}=\alpha x_i + (1-\alpha)*\frac{\beta\sum\limits_{j\in\Omega_{i}^{r}} x_j+ (1-\beta)\sum\limits_{j\in\Omega_{i}^{b}} x_j}{\beta|\Omega_{i}^{r}|+(1-\beta)|\Omega_{i}^{b}|}
\end{equation}    
where $\alpha$ represents the stubbornness level, i.e., the openness to the external influence from neighbors. The term $\frac{\beta\sum\limits_{j\in\Omega_{i}^{r}} x_j+ (1-\beta)\sum\limits_{j\in\Omega_{i}^{b}} x_j}{\beta|\Omega_{i}^{r}|+(1-\beta)|\Omega_{i}^{b}|}$ accounts for the weighted average of attractiveness for agent $i$.
\section{Results}

Many empirical studies suggest that real social networks tend to have, or at least can be approximated as a power-law distribution \cite{barabasi1999emergence}. In order to mimic the real-world circumstance, here we perform simulations of our model on scale-free networks with $N=1000$ nodes, whose degree distribution satisfies $p_k\sim k^{-3.2}$. In binary decision-making processes, we mainly focus on the stationary proportion of agents supporting one decision. Note that the proportions are almost unchanged after 100 steps (see Fig. \ref{Fig.2}). Thus, we run each simulation for 100 steps to approximate the steady state. In addition, to eliminate the fluctuation, the following simulation results are all averaged over 10 times. The robustness of our results on network scale ($N$) and the number of simulations is provided in Supplementary Information.

\subsection{Emergence of phase transition}

\begin{figure*}[!h]
\includegraphics[width=0.95\textwidth]{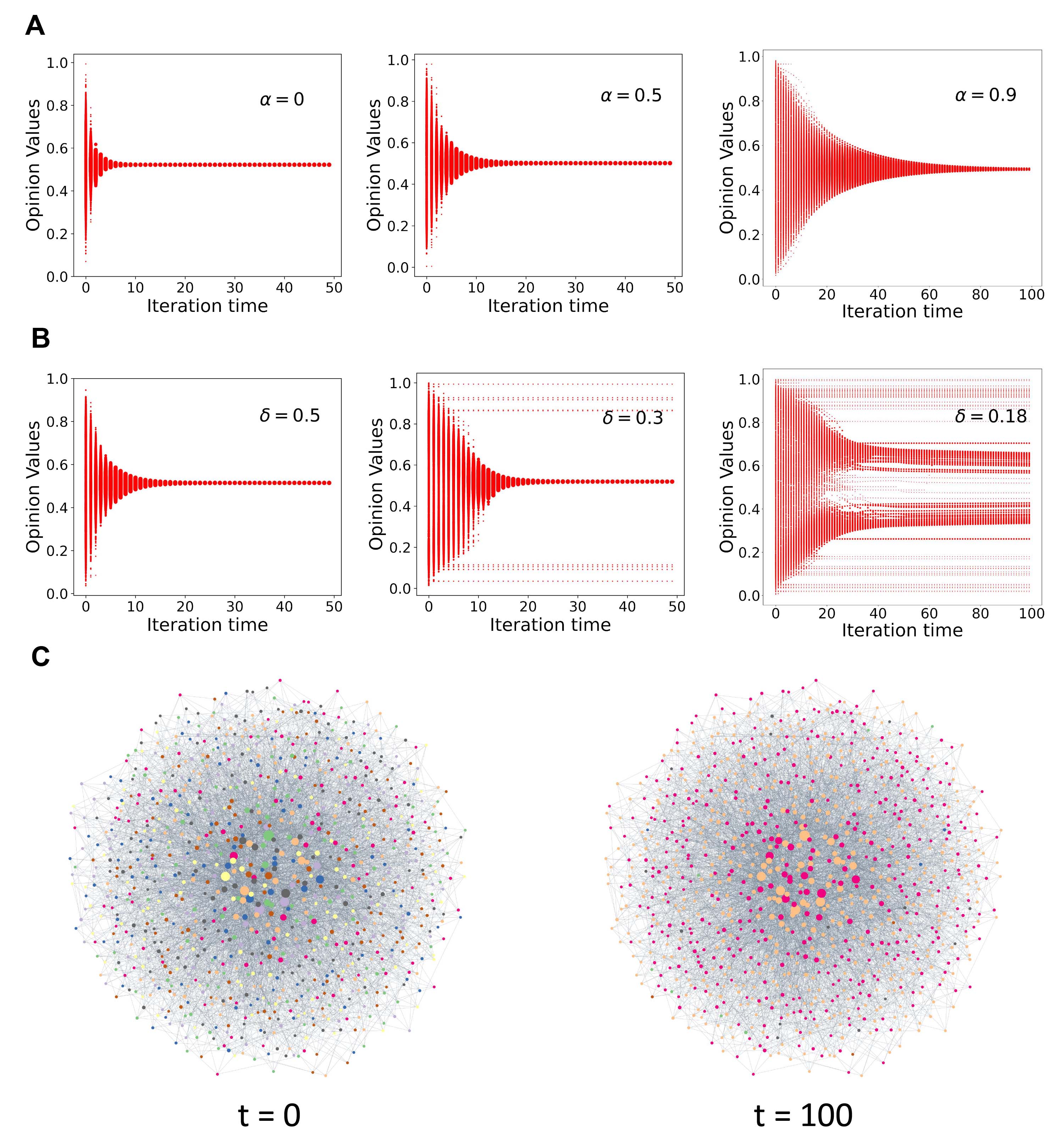}
\caption{Effects of stubbornness level and bounded confidence on system time evolutions. (A) Shown are opinion values as a function of time when $\alpha=0, 0.5$ and $0.9$, respectively. The comparisons among these subplots show that convergence time would increase with stubbornness level growing. Parameters: $\delta=0.5$, $\beta=0.5$. (B) We present time evolutions of opinion values when $\delta = 0.5, 0.3, 0.18$, respectively. For large $\delta$, the population excluding few extremists would reach consensus. For small $\delta$, there emerges two opinion clusters. Parameters: $\alpha=0.5$, $\beta=0.5$. (C) We show detailed opinion evolutions on structured networks for $\delta=0.18$, $\alpha=0.5$, $\beta=0.5$, corresponding to the third subfigure in (B).}
\label{Fig.2}
\end{figure*}

In this subsection, unless it is specified, the initial proportion of agents supporting decision red is set as 0.5 and their initial preferences (opinion values) are distributed randomly.

To begin, we explore the effect of stubbornness level ($\alpha$) by presenting how opinion values change over time under different values of $\alpha$ in Fig. \ref{Fig.2}(A). The common of all cases is that the population would finally reach consensus. The difference is that convergence time gets longer for larger $\alpha$. This adheres to our intuition that high stubbornness level could reduce the efficiency of communication in public discourse.

Then we consider the effect of bounded confidence ($\delta$). Specifically, Fig. \ref{Fig.2}(B) presents time evolutions of opinions under different values of $\delta$. When $\delta=0.5$, all the agents could almost freely interact with their neighbors, which helps the population reach consensus rapidly. For moderate $\delta$ ($\delta=0.3$), few extremists would remain initial extreme opinions while most agents reach consensus. When $\delta$ is small ($\delta=0.18$), there emerge two giant opinion clusters (Fig. \ref{Fig.2}(C)), which corresponds to the two polarized decisions in decision-making processes. Overall, the variation of bounded confidence could lead to the phase transition from the consensus to the polarization of collective decisions. Moreover, Fig. \ref{Fig.2} shows that the proportion of agents supporting one decision would become stable within 100 steps for most situations.


\begin{figure*}[!h]
\includegraphics[width=0.95\textwidth]{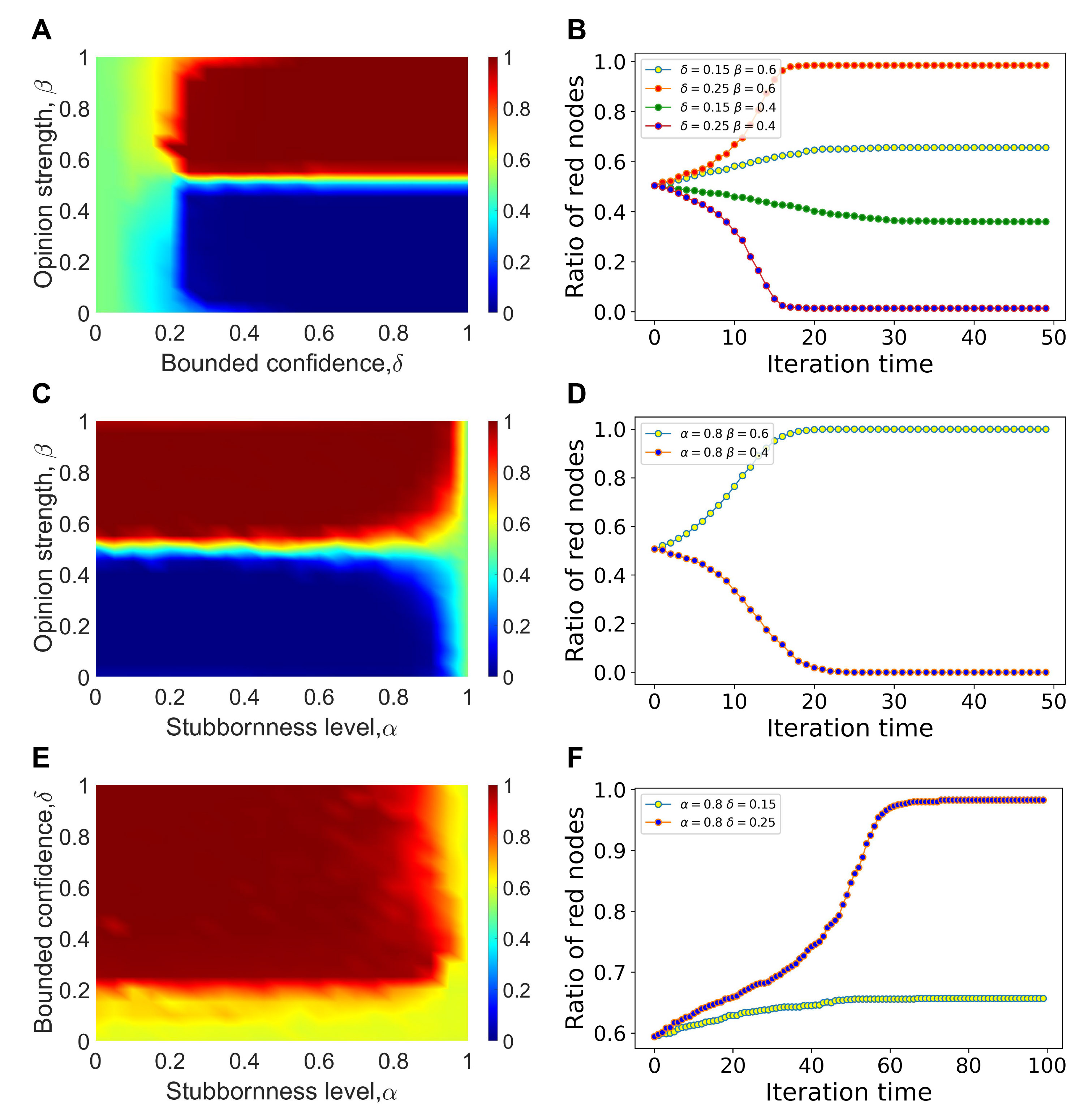}
\caption{Joint effect of different parameters. (A)(B) The joint effect of bounded confidence ($\delta$) and asymmetric opinion strength ($\beta$). (A) Shown is the phase diagram for the proportion of red supporters under different combinations of $\delta$ and $\beta$. The detailed time evolutions of some representative cases are presented in (B). (C)(D) The joint effect of asymmetric opinion strength ($\beta$) and stubbornness level ($\alpha$). (C) presents the phase diagram and (D) presents the detailed time evolutions of some cases. (E)(F) The joint effect of bounded confidence ($\delta$) and stubbornness level ($\alpha$) under asymmetric initial proportions. (E) and (F) presents the phase diagram and the time evolutions, respectively. Parameters: (A)(B) $\alpha=0.5$, $P=0.5$, (C)(D) $\delta=0.5$, $P=0.5$, (E)(F) $\beta=0.5$, $P=0.6$.}
\label{Fig.4}
\end{figure*}

In addition, we illustrate the significant effect of asymmetric opinion strength. Denote the initial proportion of agents supporting decision red as $P$. Intuitively, if initial proportions of agents supporting the two decisions are the same ($P=0.5$), the decision with higher opinion strength would finally gain more supports. This naturally arises some interesting problems: what is the threshold of asymmetric opinion strength leading to the disappearance of the disadvantaged decision? How does bounded confidence and stubbornness level influence the impact of opinion strength?

To comprehensively understand these problems, we first present phase diagrams for the stationary proportion of agents supporting decision red under different combinations of its opinion strength ($\beta$) and bounded confidence ($\delta$) in Fig. \ref{Fig.4}(A). Intriguingly, there appears a phase transition from the coexistence of both decisions to the disappearance of the disadvantaged decision at a threshold of $\delta$ (around 0.2). To be specific, when $\delta$ exceeds the threshold, a slight  advantage in opinion strength could help the decision dominate the whole population. Conversely, when $\delta$ is smaller than the threshold, both decisions coexist even if the difference in opinion strength is huge. The above critical phenomenon could be explained by the following microscopic reasons. Large bounded confidence allows agents to affect more opponents, which enhances the impact of asymmetric opinion strength. A extreme case is $\delta=1$ where agents with low-persuasive decision could freely interact with neighbors and would finally be persuaded. Small bounded confidence implies that agents with high-persuasive decision have no access to affect opponents with extreme views. This results in the survival of the disadvantaged decision. Fig. \ref{Fig.4}(B) further presents the detailed time evolutions under different $\delta$, which again illustrates the critical phenomenon caused by bounded confidence. These above two situations help provide an explanation for both the polarization in political elections and the monopoly phenomena in competitive processes of products.

In Fig. \ref{Fig.4}(C), we then explore the joint effect of asymmetric opinion strength ($\beta$) and stubbornness level ($\alpha$) by presenting phase diagram for the final proportion of red supporters. Here we set $\delta$ as a large value ($\delta=0.5$). Results show that asymmetric opinion strength would result in the domination of the high-persuasive decision under most situations excluding where $\alpha$ is extremely large. The survival of the disadvantage decision might be because large stubbornness level weakens the efficiency of communication, which causes the system to fail to reach a steady state within 100 steps. Thus we can conclude that the stubbornness level shows no significant impact when $\delta$ is large. Fig. \ref{Fig.4}(D) presents the detailed time evolutions showing how the disadvantaged decision disappears.

In addition to asymmetric opinion strength, the initial proportions also play a pivotal role in the final evolutionary results. Here we consider the situation with asymmetric initial proportions where decision red has first-move advantage ($P=0.6$). Fig. \ref{Fig.4}(E) and Fig. \ref{Fig.4}(F) presents how the bounded confidence and the stubbornness level affect the first-move advantage. We find that the first-move advantage could always maintain. In particular, when $\delta$ is larger than a threshold (around 0.2), decision with the first-move advantage could dominate the whole population. Moreover, stubbornness level shows little impact on the evolutionary results.

In summary, we highlight the phase transition from the coexistence of both decisions to the domination of a single one when the bounded confidence goes through a threshold. 

\subsection{Reversing the first-move advantage in competitive dynamics}

Through observing Fig. \ref{Fig.4}, we also find that either opinion strength or initial proportions could directly affect the competing results. Then it is natural to ask how the competition of these two asymmetric conditions affects evolutionary results. In particular, an interesting problem is on what conditions the first-move advantage could be reversed.

\begin{figure*}[!h]
\includegraphics[width=0.95\textwidth]{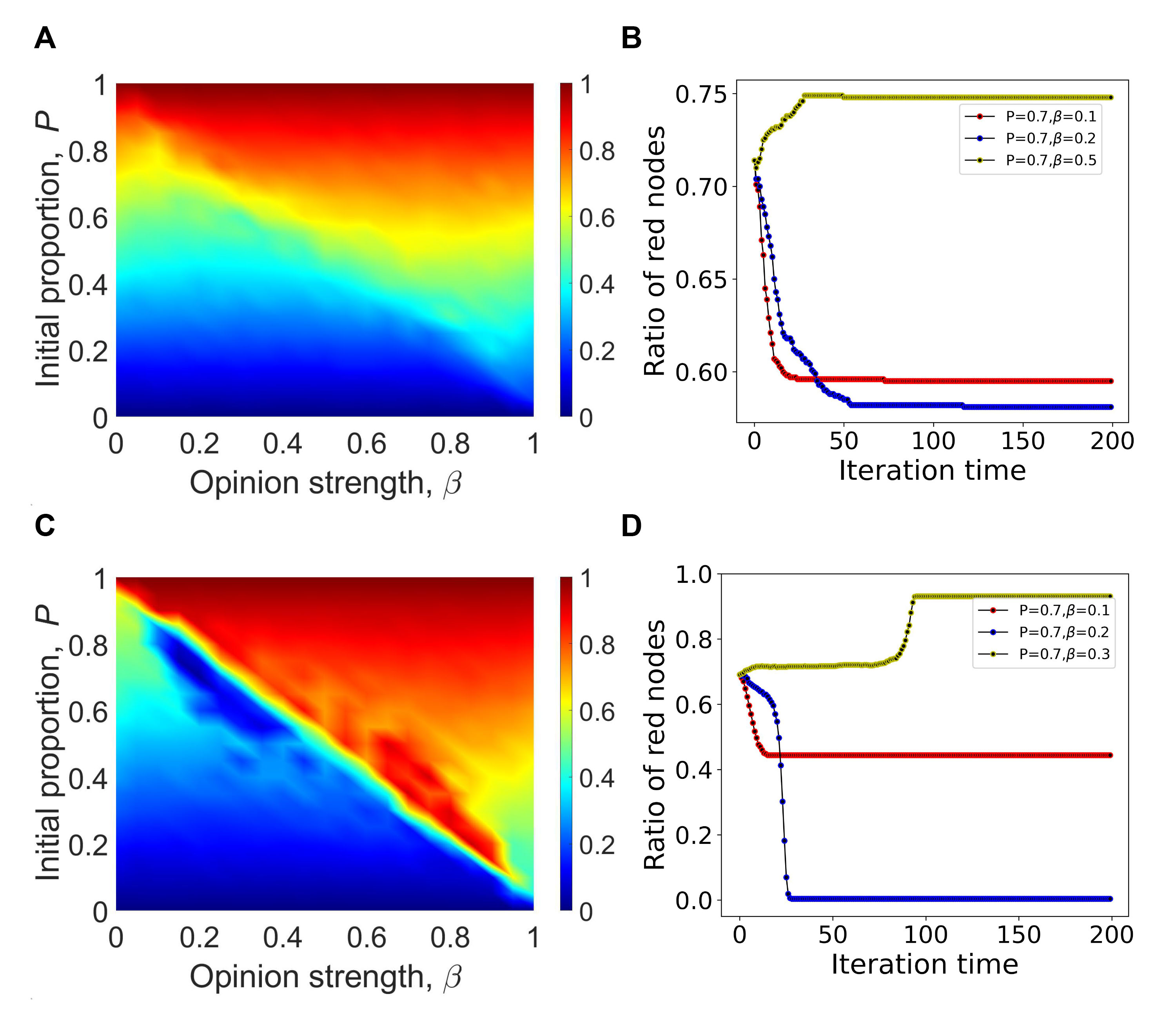}
\caption{Reversal phenomenon. We fix (A)(B) $\alpha=0.5,\delta=0.1$ and (C)(D) $\alpha=0.5,\delta=0.2$, respectively. (A)(C) Shown are phase diagrams for the final proportion of red supporters under different values of bounded confidence and opinion strength. (B)(D) Shown are the detailed time evolutions of some representative cases.}
\label{Fig.5}
\end{figure*}

We first consider the joint effect of opinion strength and initial proportions when the bounded confidence is small, i.e., $\delta=0.1$ (Fig. \ref{Fig.5}(A) and \ref{Fig.5}(B)). Even huge difference in opinion strength could not reverse the first-move advantage. The results are completely different when $\delta$ is relatively large (Fig. \ref{Fig.5}(C) and \ref{Fig.5}(D)). The reversal phenomenon caused by asymmetric opinion strength take places in most cases. The comparison indicates that reversing the first-move advantage requires the conjugate function of the bounded confidence and opinion strength. Moreover, in these figures, we find that opinion strength displays non-monotonous impacts on the evolutionary results. Specifically, the stationary proportion of red supporters first increases and then reduces with opinion strength ($\beta$) growing. This counterintuitive phenomenon might be explained from the microscopic dynamical pattern. Under extremely large $\beta$, red supporters would aggregate so rapidly that their opinions are beyond the identity scope of most moderates.       

\subsection{A case study for cascading collapse}

\begin{figure*}[!h]
\includegraphics[width=0.95\textwidth]{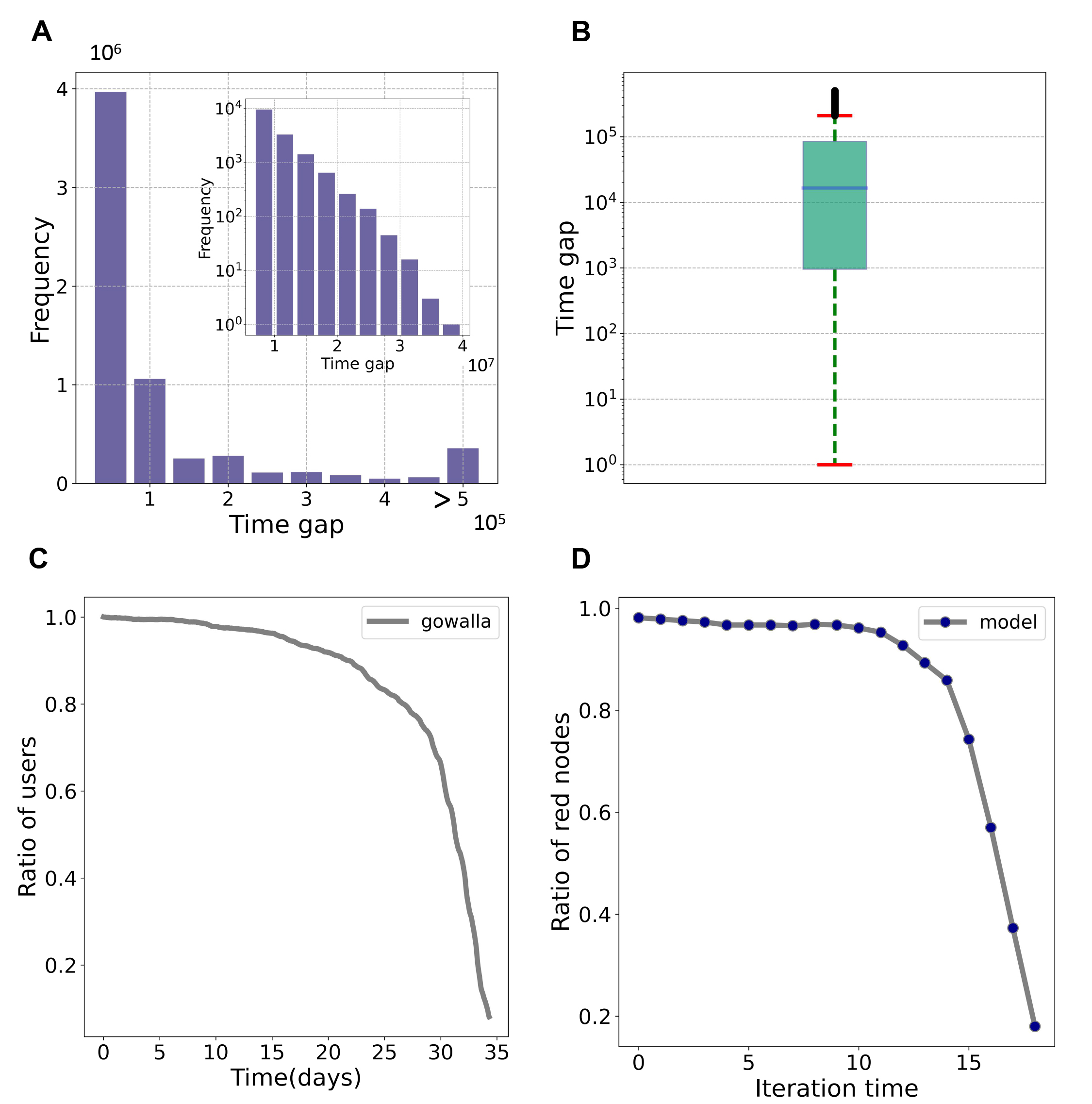}
\caption{Cascading collapse of Gowalla's failure in the competition with Foursquare. (A-B) present distributions of the time intervals between two consecutive check-ins of all users in Gowalla. Inset in (A) shows log scale of frequency, indicating that the users' check-in time intervals can be approximated to a scale-free distribution. (B) The box plot of the time intervals. (C) Cascading collapse of Gowalla's active users. If the time interval between last check-in and the ending time of the dataset is larger than a threshold, which is selected according to (B), then we assume that user is no longer active. (D) Time evolution of the proportion of red nodes in our model where $P=0.7,\alpha=0.5,\delta=0.25,\beta=0.27$. Note that we have normalized the ratio for comparison. }
\label{Fig.6}
\end{figure*}

\begin{figure*}[!h]
\includegraphics[width=0.95\textwidth]{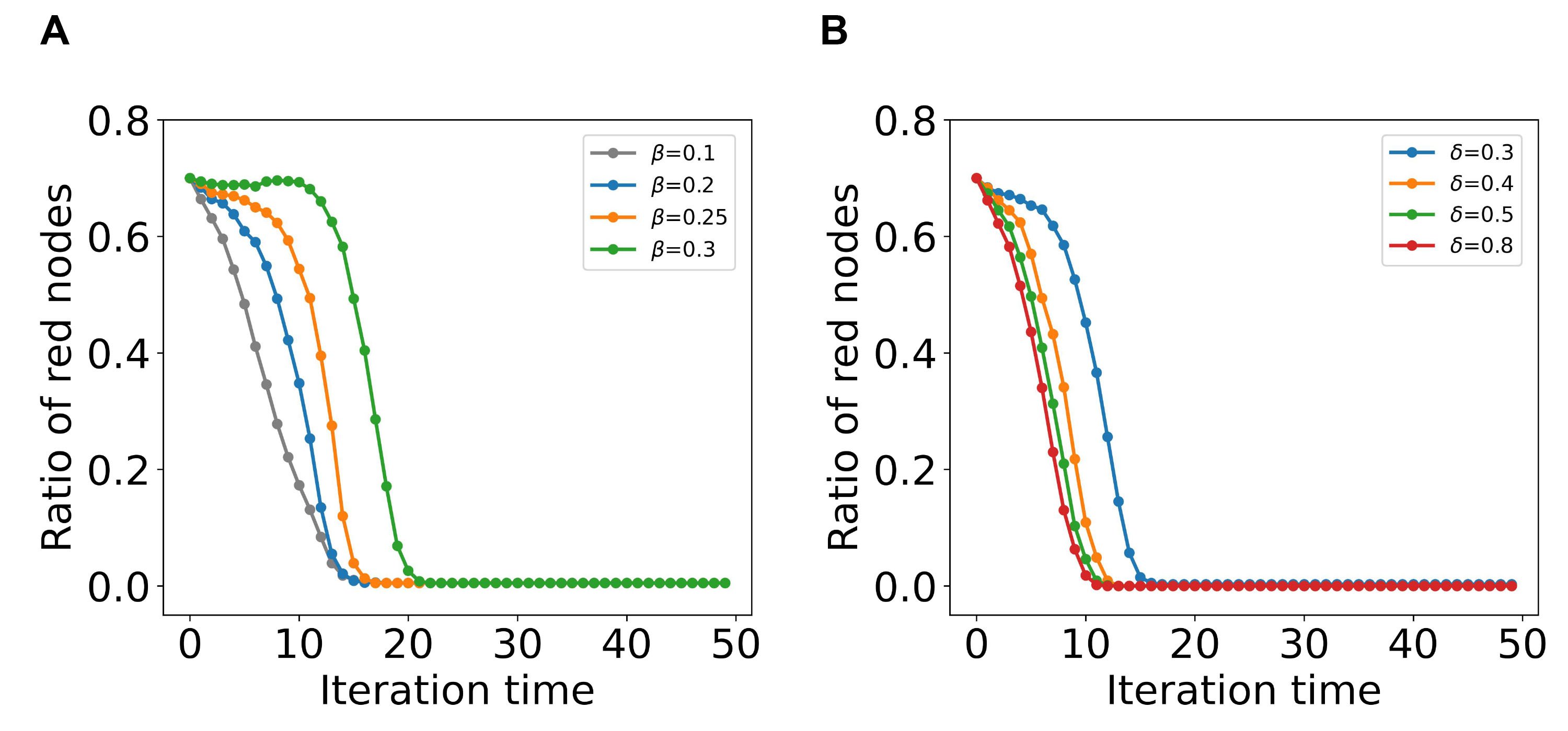}
\caption{Emergence of cascading collapse in our model. (A) The effect of opinion strength on cascading patterns during a competition. We fix $\delta=0.25$ and change $\beta=0.1, 0.2, 0.25, 0.3$, respectively. (B) The effect of bounded confidence on cascading patterns during a competition. We fix $\beta=0.3$ and change $\delta=0.3, 0.4, 0.5, 0.8$, respectively.. Other parameters: $P=0.7,\alpha=0.5$. }
\label{Fig.7}
\end{figure*}

In the brutal competition of binary decisions such as competitive ideas, behaviors or products, the cascading collapse has been widely identified as an important dynamical pattern especially for the early predictions on final evolutionary outcomes \cite{torok2017cascading,robards2012leaving,lHorincz2019collapse}. In Fig. \ref{Fig.6}, we show the emergence of cascading collapse in Gowalla's failure when mainly competing with another homogenous product Foursquare, which can be approximated to a binary decision-making process. Gowalla is a location-based social networking website, which was launched in 2007 and closed in 2012. The dataset collects $6442890$ check-in behaviors of $196591$ users over the period of Feb. 2009 - Oct. 2010 \cite{leskovec2014snap}. In order to identify the users' final state in Gowalla, either active or inactive, we use a cut-off strategy: If the time interval between last check-in and the ending time of the dataset for a certain user is larger than a threshold, we assume that user is no longer active; Otherwise, we assume the user is still active. To determine a reasonable cut-off threshold, we first present the distribution of all time intervals between two consecutive check-ins of all users (Fig. \ref{Fig.6}A). The subfigure embedded in Fig. \ref{Fig.6}A further shows that the users' check-in time intervals can be approximated to a power-law distribution. In other words, the majority of the time intervals are relatively concentrated, which are within $100,000$ seconds, about $1.2$ days. Therefore, we can safely choose the largest whisker of the box plot in Fig. \ref{Fig.6}B, which is $210137$ seconds, about $2.4$ days, as the critical point that divides the users into active or inactive states. Beginning from the maximal value of active users, we show time evolution of the ratio of Gowalla's active users (Fig. \ref{Fig.6}C), which decreases slowly first and then drops sharply, indicating the emergence of a typical cascading collapse. Notably, in Fig. \ref{Fig.6}D, we qualitatively reproduce such dynamical behavior using our model, where $P=0.7,\alpha=0.5,\delta=0.25,\beta=0.27$. Each iteration time step in our model corresponds to approximately two days in real world. 

Furthermore, we explore the detailed impacts of opinion strength and bounded confidence on cascading patterns in Fig. \ref{Fig.7}A and Fig. \ref{Fig.7}B, respectively. Results show that a moderate level of opinion strength (in this case, $\beta \in [0.25, 0.3]$) and a moderate bounded confidence are most likely to contribute to the emergence of cascading collapse that is close to reality. And a strong opinion strength or a large bounded confidence may both lead to a direct sharp descent in the failure of binary decision-making dynamics. 

\subsection{Effects of individual cognitive heterogeneity}
By far, our results are constructed on the assumption of population cognitive homogeneity where all people have the same cognitive scope (i.e., bounded confidence $\delta$). In this section, we further relax this homogenous assumption to a more realistic heterogeneous situation where the population is consist of two classes of individuals: those who are less open-minded ($\delta_1=0.1$) and those who are open-minded ($\delta_2=0.3$). Denote $\rho$ as the fraction of less open-minded people. 

\begin{figure*}[!h]
\includegraphics[width=0.95\textwidth]{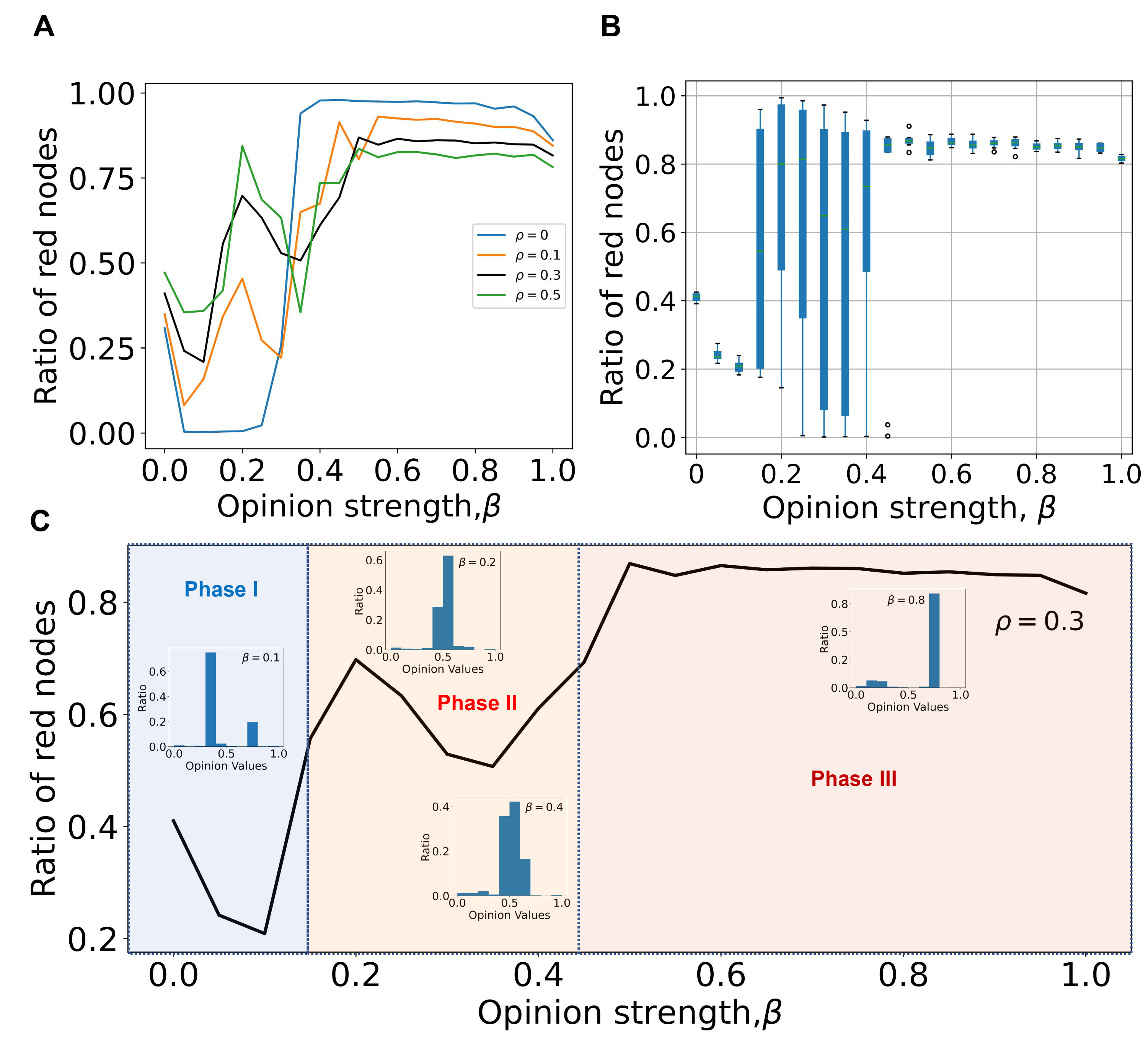}
\caption{Emergence of indeterministic oscillation under the influence of individual cognitive heterogeneity. (A) How the individual cognitive heterogeneity influences binary decision-making dynamics. For simplicity, we consider a simple heterogeneous situation where the population is consist of two classes of individuals: those who are less open-minded ($\delta_1=0.1$) and those who are open-minded ($\delta_2=0.3$). Denote $\rho$ as the fraction of less open-minded people. We present the ratio of red nodes as a function of opinion strength and change $\rho=0, 0.1, 0.3, 0.5$, respectively. Simulation results are averaged over $10$ independent runs. (B) Box plot of $10$ times of simulations for a heterogeneous scenario where $\rho=0.3$, corresponding to the black curve in (A). Results clearly show the emergence of an indeterministic oscillation region. (C) Phase transitions and opinion distributions under the existence of individual cognitive heterogeneity. Other parameters for (A)-(C): $P=0.7,\alpha=0.5$. }
\label{Fig.8}
\end{figure*}

In Fig. \ref{Fig.8}A, we show how the individual cognitive heterogeneity influences binary decision-making dynamics. The ratio of red nodes is present as a function of opinion strength ($\beta$) and we change $\rho=0.1, 0.3, 0.5$, respectively. For comparison, we also provide a homogeneous situation where $\rho=0$, indicating that all people are relatively open-minded with $\delta=\delta_2=0.3$. Other parameters are fixed as $P=0.7,\alpha=0.5$. Overall, all heterogeneous scenarios share the same changing pattern: as $\beta$ increases, the ratio of red nodes first decreases, then rises and experiences a period of irregular oscillation, and finally evolves to a stable high plain. When $\rho$ becomes larger, the increase of less open-minded individuals compromises the substantial changes caused by open-minded individuals, making it more difficult for both reversing the first-move advantage (see region $\beta \in [0, 0.1]$) and amplifying the first-move advantage (see region $\beta \in [0.5, 1]$). 

Of particular interest, in Fig. \ref{Fig.8}B, we provide box plot to explore the detailed dynamical patterns of a typical heterogeneous circumstance where $\rho=0.3$. Interestingly, results clearly show the emergence of an indeterministic oscillation region ($\beta \in [0.15, 0.45]$) where the final ratio of red nodes covers a great range in different simulations. This indicates that the existence of individual cognitive heterogeneity may bring about randomness and unpredictability in binary decision-making process. The final outcome of the competitive dynamics within this area can be any scenario: either Red/Blue becomes dominant, or Red and Blue coexist with any mixing ratio. 
 
To better understand this non-intuitive phenomenon, in Fig. \ref{Fig.8}C, we further show phase transitions and opinion distributions under the existence of individual cognitive heterogeneity. The curve is divided into three phases according to the results in Fig. \ref{Fig.8}B. In phase \uppercase\expandafter{\romannumeral1}, we observe the backfire effect of the extremely strong opinion strength, which is consistence with the homogeneous situation (Fig. \ref{Fig.5}C). Phase \uppercase\expandafter{\romannumeral2} corresponds to the indeterministic oscillation region. And phase \uppercase\expandafter{\romannumeral3} represents the stable states where the first-moved advantage is preserved and slightly enlarged. The embedded subfigures in each area show how opinion distribution changes along with the occurrence of phase transitions. We find that as $\beta$ increases, not surprisingly, there is a general trend for population opinions to move from left to right, namely, from adopting Blue decisions to adopting Red decisions. However, it is worthy of note that in phase \uppercase\expandafter{\romannumeral2}, the population opinions eventually surround the moderate positions (at around $0.5$). In other words, a small disturbance on opinion dynamics may lead to substantial changes on binary decisions, which accounts for the indeterministic oscillation at the macroscopic level. Under such circumstance, intuitively, the final result may largely depend on the initial distribution of hubs' opinions.

\section{Conclusion and discussion}

Over the last few years, the ways of information transmission and opinion formation have been radically reshaped with the rise of online social networks \cite{wang2017promoting}. On one hand, information spreads faster and more widely, making it more convenient for individuals to obtain ideas and viewpoints. On the other hand, information can also be reprocessed, selectively diffused and easily manipulated on various large-scale social networks \cite{aral2019protecting}. An important consequence of these changes might be the appearance of the increasing polarization on many social issues \cite{macy2019opinion}, which further addresses the importance of understanding and predicting the dynamics of binary decision-making process. In the context of the ongoing COVID-19 pandemic, the collective binary decision-making outcome of individual attitudes toward vaccination may even affect the fate of human societies \cite{wang2020eco}. 

In this work, we propose an agent-based model of competitive opinion evolution to characterize the group binary decision-making dynamics. The main contribution of our frame is the incorporation of opinion strength, which describes the asymmetry power of binary decisions on social influence process. Importantly, through this simple mechanism, we are able to explore the coupled interactions between binary behaviors and continuous beliefs, while most previous works study these two dynamical processes separately. Our model identifies two important physical phenomena. The first is the phase transition from co-existence (polarization) to winner-takes-all (consensus) state, which is largely determined by the bounded confidence. The second is the cascading collapse in competitive dynamics, which qualitatively reproduces the actual dynamical pattern observed in the failure of Gowalla when competing with a homogenous product Foursquare. 

Particularly, we are interested in exploring the detailed conditions of reversing the initial supportive advantage, which can also be understood as the `first-move advantage' in a competition. We illustrate that the reversal can only happen under the combined influence of bounded confidence and opinion strength. More intriguingly, we find that an overly strong opinion strength may be harmful for the competitive advantage, as opinions of those who support the favorable decision may quickly be far away from opinions of their neighbors who still hold opposite decision. Due to the existence of bounded confidence, the impact of social influence hardly works under this circumstance and the `persuasion' process becomes less. This counter-intuitive finding is actually in line with a latest advance which suggests that extreme political advertising can hurt campaign efforts \cite{wang2020public}.

The importance of population cognitive scope (bounded confidence) cannot be overstated. To better simulate the real-world situation, we also consider heterogeneous populations where individuals are assigned with different cognitive scope values, either open-minded (a relatively large bounded confidence) or less open-minded (a relatively small bounded confidence). Surprisingly, we observe the emergence of an indeterministic oscillation region where any possible result of the competition could happen, including one choice wins all and both choices coexist. We conclude that individual cognitive heterogeneity may result in randomness and unpredictability in binary decision-making dynamics.
 
Our model reveals how the rich dynamical behaviors of binary-decision making in real world can be understood as the results of belief-behavior interactions via certain elements of competitive opinion evolution. Therefore, our findings may provide valuable insights toward competitive dynamics in many different fields, including the increasing competition between industry products, the public health efforts, and the political campaigns. Further studies may incorporate the heterogeneity of population and provide more precise models based on larger datasets. 

\ack{This work is supported by Program of National Natural Science Foundation of China Grant No. 11871004, 11922102, and National Key Research and Development Program of China Grant No. 2018AAA0101100.}


\providecommand{\newblock}{}

\end{document}


\title{Supplementary Material}
\section{Robustness of simulation results}
\begin{figure*}[!h]
\includegraphics[width=0.95\textwidth]{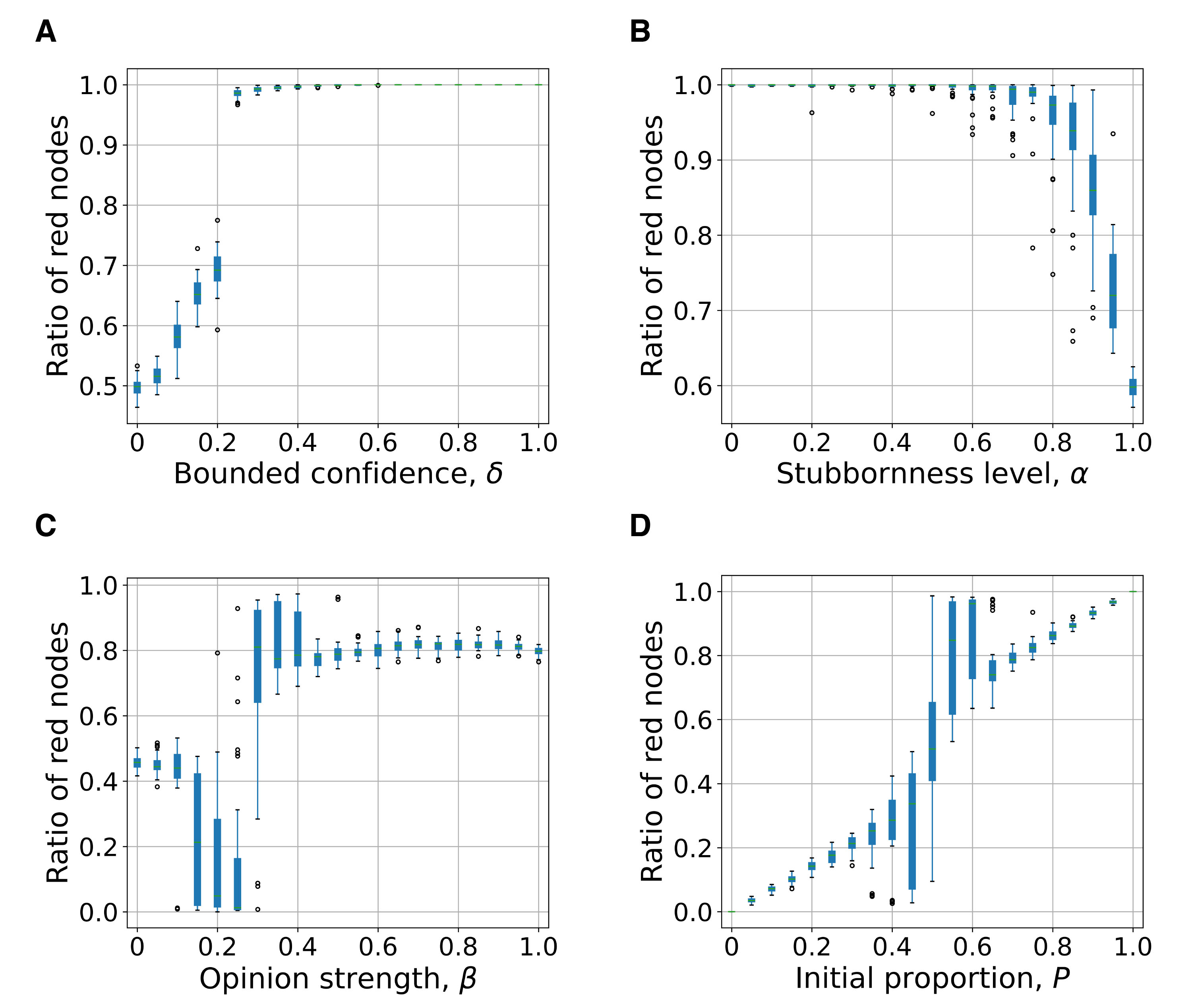}
\caption{Robustness of simulation results for all parameters. All simulations are independently performed for 30 times. Parameters: (A) $\alpha=0.5$, $\beta=0.8$, $P=0.5$. (B) $\beta=0.5$, $\delta=0.5$, $P=0.6$. (C) $\alpha=0.5$, $\delta=0.2$, $P=0.7$. (D) $\alpha=0.5$, $\beta=0.5$, $\delta=0.2$.}
\label{Fig.1}
\end{figure*}

In this section, all simulations are independently performed for 30 times in order to explore the robustness of our results. In Fig. \ref{Fig.1}, we present box plots of the stable states as a function of $\delta$, $\alpha$, $\beta$ and $P$, respectively. Results show that our simulations are robust to all parameters in most cases, while a moderate oscillation would occur when the system approaches to the tipping points where phase transitions emerge. Note that the stochastic nature is inevitable within the parameter regions near the tipping points.

\section{Effect of the system size}
\begin{figure*}[!h]
\includegraphics[width=0.95\textwidth]{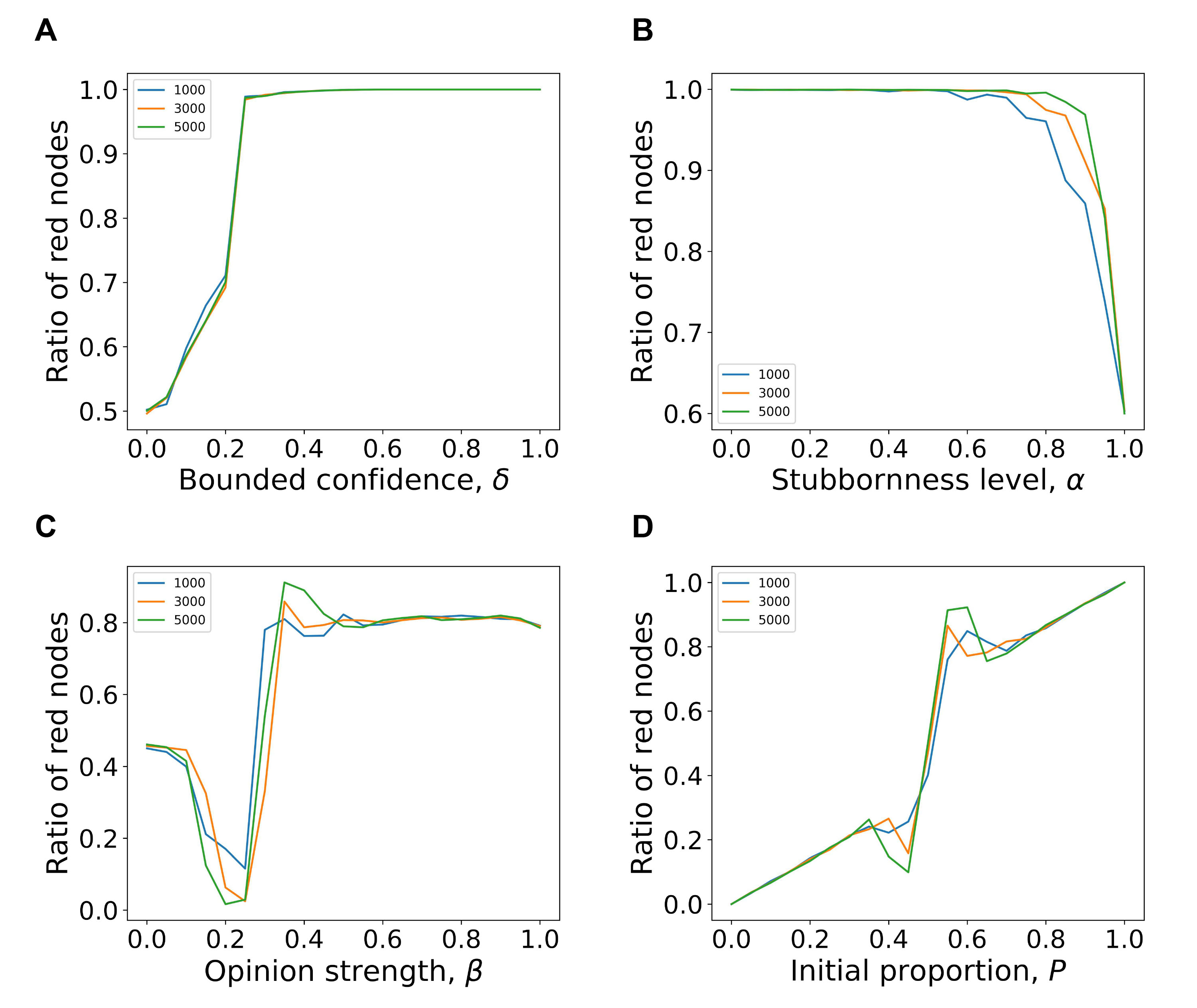}
\caption{Effect of the system size. The results are averaged over 10 independent runs under different system sizes: $N=1000$, $N=3000$, $N=5000$. Parameters: (A) $\alpha=0.5$, $\beta=0.8$, $P=0.5$. (B) $\beta=0.5$, $\delta=0.5$, $P=0.6$. (C) $\alpha=0.5$, $\delta=0.2$, $P=0.7$. (D) $\alpha=0.5$, $\beta=0.5$, $\delta=0.2$.}
\label{Fig.2}
\end{figure*}

Fig. \ref{Fig.2} shows detailed influence of the system size. All simulations are averaged over 10 independent runs. We find that under all circumstances, the final ratio of red nodes has no significant change when the system size changes ($N=1000, 3000, 5000$, respectively). We conclude that our results are robust when $N$ is relatively large.

